\documentclass[final,5p,times,twocolumn,authoryear]{elsarticle}

\usepackage{amssymb}
\usepackage{booktabs}
\usepackage{adjustbox}
\usepackage{siunitx}
\usepackage{lscape}
\usepackage{multirow}
\usepackage{tikz}
\usetikzlibrary{positioning,calc}
\usepackage{colortbl}
\usepackage{array}
\usepackage{svg}
\usepackage{graphicx} 
\usepackage[unicode,breaklinks=true]{hyperref}
\hypersetup{
	colorlinks=true,
   	linkcolor=blue,
   	citecolor=blue,
    filecolor=blue,
   	urlcolor=blue}
\usepackage{colortbl}
\usepackage{hhline}
\usepackage{pdflscape}
\usepackage{caption} 
\usepackage{graphicx,calc}
\usepackage{makecell}

\newcolumntype{L}[1]{>{\raggedright\let\newline\\\arraybackslash\hspace{0pt}}m{#1}}
\newcolumntype{C}[1]{>{\centering\let\newline\\\arraybackslash\hspace{0pt}}m{#1}}
\newcolumntype{R}[1]{>{\raggedleft\let\newline\\\arraybackslash\hspace{0pt}}m{#1}}

\usepackage{soul}
\usepackage{color}
    \definecolor{celadon}{rgb}{0.67, 0.88, 0.69}
    \definecolor{flamingopink}{rgb}{0.99, 0.56, 0.67}
    \definecolor{lovelygreen}{rgb}{0.54, 0.90, 0.60}

\usepackage[figuresright]{rotating}
\usepackage{pbox}
\usepackage{makecell}
\usepackage{appendix}

\usepackage{footnote}
\usepackage{tablefootnote}

\usepackage{graphicx}
\usepackage{tabularray}
\usepackage{xcolor}

\definecolor{gray7}{HTML}{BFBFBF}
\definecolor{gray9}{HTML}{E5E5E5}

\usepackage{tabularray}

\makesavenoteenv{tabular}
\makesavenoteenv{table}

\journal{DFRWS EU 2027}

\begin{document}
\emergencystretch 3em

\begin{frontmatter}



\title{Federated Forensics: Jurisdictional constraints as a design driver for multi-agency federated learning}


\author[label1]{Aruna Withanage\corref{cor1}}
\ead{aruna.withanage@uni-a.de}
\cortext[cor1]{Corresponding author}
\author[label1,label2]{Jessica Herrmann}
\ead{jessica.herrmann@uni-a.de}
\author[label3]{Hudan Studiawan}
\ead{hudan@its.ac.id}
\author[label1]{Frank Breitinger}
\ead{frank.breitinger@uni-a.de}

\affiliation[label1]{organization={Chair for Cybersecurity, Institute of Computer Science, University of Augsburg},
             city={Augsburg},
            country={Germany}}

\affiliation[label2]{organization={Präsidium Technik, Logistik und Service der Polizei Baden-Württemberg},
             city={Stuttgart}, 
             country={Germany}}
\affiliation[label3]{organization={Department of Informatics, Institut Teknologi Sepuluh Nopember},
             % city={}, 
             country={Indonesia}}

\begin{abstract}

Digital forensic investigations increasingly cross borders, and law enforcement agencies need to build analytical models together without centralizing their data. Federated Learning (FL) keeps data local, yet neither FL nor its frameworks enforce the jurisdictional, governance, and accountability rules that multi-agency forensic work requires.
This paper takes a design-first approach to that gap. First, we derive seven requirements that a federated learning system must meet in a forensic setting.  
Second, we propose \emph{Federated Forensics}, a three-layer reference architecture whose middle layer, unlike in existing hierarchical FL designs, sits at a jurisdictional border and can use a model update locally while blocking it from crossing that border. Third, we build a prototype of the architecture and test it on a malware-family classification task.
In a prototype evaluation, the policy gate blocked every restricted update at the border with zero violations, and restricted-class accuracy rose from 0\% under today's compliant alternative to 93.0\%, only 4.4 points below ignoring the restriction.

\end{abstract}


\end{frontmatter}



\section{Introduction}
\label{sec:introduction}

Law enforcement agencies (LEAs) rely on digital evidence, and many investigations now cross national borders \citep{hargreaves2024dfpulse}. Legal instruments already allow agencies to exchange evidence across borders \citep{eio2014,eevidence2023, reviewOfreview}. Training a machine learning model together is a different matter: a model trained on data from several agencies may reveal patterns that no single agency can see, but the training data is sensitive and often cannot be collected in one place \citep{scanlon2023chatgpt,breitinger2023sharing}. Federated Learning (FL) addresses this technically, since each agency keeps its data and only model updates are shared \citep{mcmahan2017communication}. However, existing FL frameworks were not built for forensic work: auditability, governance across organizational levels, explainability \citep{solanke2022explainable}, and evidentiary accountability are rarely design goals \citep{riedel2024comparative}.

Some legal requirements can be met through system design, for example by minimizing data, logging processing, and restricting access. Others, such as the legal basis, purpose limitation, and lawful further use, require legal assessment \citep{directive2016680,TeMolder2023}. Cross-border cooperation adds a specific question. In FL, evidence is always analyzed locally, so the question is whether information derived from it may contribute to a model that other jurisdictions receive.

Here, a flat FL topology reaches a limit. It can comply with a restriction by leaving the restricted material out of training. But it produces only one shared model, so the material is then also missing from the model used inside the jurisdiction, where its use is lawful and often most valuable. A flat topology can exclude restricted material, but it cannot scope where its influence travels. Scoping requires an aggregation point at the border of the jurisdiction, which is a property of the topology, not a setting \citep{9148862}.

This paper therefore asks: \emph{which federated learning topologies allow a jurisdictional restriction to be enforced at the point where it legally applies, and what does enforcing it cost in model quality?}

We make three contributions. First, we derive seven requirements (R1 to R7) for FL in forensic settings, state what the forensic context adds beyond other regulated domains, identify a conflict between two of them, and assess the open-source FL ecosystem against them. Second, we propose \emph{Federated Forensics}, a three-layer reference architecture whose middle layer sits at the border of legal competence and can withhold a contribution from further aggregation. Third, we evaluate a prototype in a controlled experiment: the gate blocked every restricted update, and accuracy on the restricted classes rose from 0\% under the compliant flat alternative to 93.0\%, 4.4 points below ignoring the restriction.

Two delimitations apply. First, our results are intelligence, not evidence: a model update loses its link to specific items of raw evidence once it leaves the agency, so a finding can direct evidence collection under chain-of-custody procedures but never replaces it. Second, we do not design the audit mechanism. The architecture fixes where audit records must be written, but their format, verification, and conflict with differential privacy (Sec.~\ref{subsec:req_tensions}) are treated in separate work.

Sec.~\ref{Federated Learning and Digital Forensics: Context and Gaps} positions the work, Sec.~\ref{Forensic Specific Requirements for Federated Learning} derives the requirements, Sec.~\ref{Federated Forensics: A Reference Architecture for Law Enforcement Collaboration} presents the architecture, Sec.~\ref{Prototype Implementation} evaluates it, and Secs.~\ref{sec:limitations} and~\ref{sec:conclusion} discuss limitations and conclude.

\section{Federated learning and digital forensics: Context and gaps}
\label{Federated Learning and Digital Forensics: Context and Gaps}

This section reviews current federated learning and identifies which of its assumptions fail in law enforcement and matter for the architecture.

\subsection{Federated learning in privacy-sensitive domains}

In FL, participants share model updates instead of data, and an aggregator combines them into a global model. Raw data stays local, so FL fits domains where regulation or ethics prevent central collection. 
In healthcare, ten institutions jointly trained models that reached 99\% of the performance of a centrally trained model and generalized to external institutions \citep{sheller2020federated}. Similar deployments exist in finance \citep{FL_bank} and in cybersecurity, notably intrusion and IoT malware detection \citep{hernandez2025intrusion,jsan14040078}.
The standard way to combine updates is Federated Averaging (FedAvg), which takes a weighted average of them \citep{mcmahan2017communication}. FedAvg assumes that all participants behave honestly. Otherwise, a single manipulated update can distort the shared model, so robust methods discard updates that deviate too far from the rest \citep{NIPS2017_Krum,FedTrimmedAvg}.
Further work deals with participants whose data differs strongly from one another, with reducing how much is sent per round \citep{konevcny2016federated}, and with hiding individual updates from the aggregator.
Consequently, one can say that FL is technically mature, but it rests on assumptions. Most research assumes cooperative participants, homogeneous governance, and evaluation aimed at model performance \citep{riedel2024comparative}, scalability, and privacy. Organizational accountability, legal admissibility, evidentiary traceability, and procedural transparency are often outside the scope of framework design. This follows from the settings FL was built for, with assumed trust, cooperative data owners, and flat governance, none of which holds in multi-agency law enforcement.

\subsection{Hierarchical federated learning}
\label{subsec:Hierarchical Federated Learning}

Multi-tier federated learning is well established. \cite{9148862} introduced the client-edge-cloud design, in which edge servers aggregate before a cloud server produces the global model. Related designs distribute aggregation across heterogeneous cellular networks \citep{abad2020hierarchical} or cluster clients by update similarity to reduce the effect of non-IID data \citep{briggs2020hierarchical}.
In all of these designs, the middle tier exists for engineering reasons. Edge aggregation reduces uplink traffic to the cloud, exploits locality and latency, lowers the frequency of expensive global synchronization, and improves convergence under heterogeneous data. The intermediate aggregator is therefore a performance optimization. It is assumed to be cooperative; it is not an authority, and it forwards whatever it has aggregated.
Our topology is therefore not novel; what differs is why the middle tier exists and what it must therefore be able to do. If the middle tier is placed where legal authority over a jurisdiction sits, it must be able to restrict, scope, or withhold what it forwards, and it must be able to justify that decision afterwards. Neither capability has a counterpart in designs built for communication efficiency, because there an intermediate aggregator never has a reason to refuse. We develop this in Sec.~\ref{Federated Forensics: A Reference Architecture for Law Enforcement Collaboration} and evaluate it in Sec.~\ref{subsec:empirical_results}.

\subsection{Federated learning in digital forensics and law enforcement}

Digital investigations differ from other FL domains. Law enforcement agencies (LEAs) must preserve evidence integrity, maintain the chain-of-custody, and keep analytical processes defensible in court. Organizational hierarchies, jurisdictional boundaries, and legal mandates shape their workflows.

FL has been proposed for forensic-adjacent tasks, but only on research data.
\cite{Parra2022InterpretableFT} presented an interpretable federated transformer for threat detection and triage in cloud environments, and \cite{10.1145/3579375.3579387} proposed an FL framework for IoT forensics analysis, evaluated on IoT attack data and silent on FL process governance and auditability. \cite{dangi2025lift} introduced LIFT for live memory forensics and real-time malware detection, with little attention to explainability, legal admissibility, or multi-agency governance. None of these works show FL used by LEAs on evidentiary material in an investigation.

A second line of work helps law enforcement and judicial authorities access cross-border electronic evidence, for example, Europol and the SIRIUS\footnote{\url{https://www.europol.europa.eu/how-we-work/sirius-project}} project. These initiatives support lawful access, provide standardized cooperation guidelines, and enable knowledge sharing. They concern data access and exchange rather than collaborative model training.

Both lines come close without covering our problem. Forensic FL work targets single tasks and assumes equal, cooperative participants; evidence-exchange initiatives govern access to data, not joint model training. In both, the law is something to comply with afterwards rather than something that shapes the design. What is missing is an FL architecture in which jurisdictional rules are part of training itself.

\section{Forensic-specific requirements for federated learning}
\label{Forensic Specific Requirements for Federated Learning}

This section derives the requirements that a federated learning system must satisfy in a law enforcement setting and examines the tensions between them.

\subsection{Rationale and scope}

The requirements give a shared vocabulary for FL in law enforcement, make explicit what framework design usually leaves out, and inform the architecture in Sec.~\ref{Federated Forensics: A Reference Architecture for Law Enforcement Collaboration}. They are neither a checklist nor a scoring system.
Several of the requirements below also appear in FL for healthcare and finance. Privacy preservation and auditability, in particular, are common to any regulated domain, so we do not claim the individual requirements as novel. Our contribution is threefold. We consolidate them into one set for the forensic setting, we state what the forensic setting adds to each of them (Table~\ref{tab:domain_contrast}), and we show that two of them conflict under conditions that do not arise in the comparison domains (Sec.~\ref{subsec:req_tensions}).

\subsection{Methodology}
Requirements were derived from prior work in three areas:

\textit{FL frameworks and mechanisms.} Comparative studies of FL systems \citep{riedel2024comparative}, foundational work on FL architectures \citep{bonawitz2019towards, kairouz2021advances}, privacy-preserving mechanisms including secure aggregation \citep{secureAggration} and differential privacy \citep{mcmahan2018learning, geyer2017differentially}, and communication-efficient protocols \citep{mcmahan2017communication}. This gave us the core architectural, privacy, and communication design principles.

\textit{Regulated domains.} FL deployments in healthcare \citep{sheller2020federated, rieke2020future, xu2021federated} and finance \citep{FL_bank, yang2019federated}. We chose these two because, among non-LEA deployments, they are close to the LEA context in three respects: strict regulatory oversight, collaboration across organizational boundaries, and high cost of errors. Mobile analytics and IoT were excluded because their FL deployments are mostly cross-device, with many devices under one operator, whereas multi-agency collaboration is cross-silo between independent organizations \citep{kairouz2021advances}.
This literature informed the requirements on governance, auditability, and compliance with sector regulations such as GDPR \citep{voigt2017eu} and HIPAA \citep{annas2003hipaa}.

\textit{Law enforcement and digital forensics.} We looked at empirical studies of investigative workflows and system requirements \citep{SKIPANES2026100805}, knowledge management in policing \citep{GOTTSCHALK201094}, hybrid intelligence systems that use open-source intelligence \citep{Maio2025HybridOSINT}, forensic process models \citep{carrier2003defining, palmer2001road}, chain-of-custody \citep{casey2011}, and forensic readiness \citep{rowlingson2004ten}. We supplemented this literature with direct practitioner input: one of the authors works for a law enforcement agency and contributed first-hand knowledge of investigative workflows, organizational structures, and operational constraints. This helped us interpret and apply the literature to real law-enforcement settings.

We coded recurring concepts, kept those appearing in more than one source, and grouped them into requirements, excluding those mainstream frameworks already handle well, such as convergence or communication efficiency.

\subsection{Core requirements}

The seven requirements (R1 to R7) are unranked, because their relative importance depends on the investigative context and the legal environment. The list is not exhaustive. It is scoped to requirements that existing FL frameworks either do not address or address insufficiently for forensic use.

\textbf{R1.} Privacy Preservation and Data Locality:
Raw data must stay under the control of the originating agency and must not be exposed to unauthorized parties. This includes jurisdictional boundaries, purpose limitation, and data minimization. Under Article~5(1)(b) GDPR and Directive (EU) 2016/680, purpose limitation binds at the level of the individual investigation \citep{gdpr_article5,directive2016680}.
R1 is often equated with the use of Secure Aggregation and Differential Privacy. In this setting, that equation is wrong. The two mechanisms address different threats, and neither is needed everywhere. Differential Privacy protects data subjects against inference by the aggregator, so its motivation depends on whether the aggregator has a lawful claim to the underlying data. Secure Aggregation hides a participant's contribution from the other participants and from the aggregator, so its motivation is operational rather than data-protective. In the deployments that FL practice comes from, the aggregator is a commercial operator with no legal relationship to the data subjects, and the first threat is acute. A regional hub that aggregates updates from its own subordinate agencies is in a different position. What R1 demands is therefore not a fixed set of mechanisms, but a decision, taken per layer, about which protection is owed to whom. Sec.~\ref{subsec:threat_model} takes that decision for the proposed architecture.

\textbf{R2.} Auditability and Traceability:
Forensic processes need end-to-end auditability. The system must record, reconstruct, and verify the training activity, including model updates, aggregation steps, and who participated. 
The standard is not internal reproducibility but external examinability. The relevant ISO standards give guidance rather than binding requirements, but the expectation is clear. ISO/IEC 27041 covers the assurance that an investigative method is suitable for its purpose and that its implementation can be shown to meet that purpose; ISO/IEC 27042 covers the analysis and interpretation of digital evidence, and ISO/IEC 27043 sets out the principles and processes that an investigation should follow and document across its lifecycle. In court, the record may be challenged by other parties, and it may need to remain verifiable long after the underlying analysis was conducted.

\textbf{R3.} Governance and Organizational Control:
LEA collaboration follows hierarchical and inter-agency governance structures. The system must support explicit governance mechanisms that define roles, responsibilities, and decision-making authority, including the management of participation, the control of model dissemination, and the enforcement of organizational policy. A key difference from consortium-based FL in other domains is that participation may be constrained by statutory competence rather than being determined solely by contractual agreement. An agency may be legally unable to contribute to a given aggregate, however willing it is. R3 is the requirement this paper concentrates on: the architecture in Sec.~\ref{Federated Forensics: A Reference Architecture for Law Enforcement Collaboration} is derived from it, and the policy gate implements it.

\textbf{R4.} Explainability and Interpretability:
Models may inform investigative decisions and, indirectly, judicial proceedings. Investigators must be able to reason about model behavior, limitations, and error sources, since outputs that cannot be interpreted have limited operational and legal use \citep{solanke2022explainable,michelet2026gptoss}. The required level of explainability may depend on how the output is used. Where outputs serve only as investigative intelligence and the resulting evidence is verified independently, it remains unclear to what extent the model itself must be explainable to a court. Where outputs feed into a legal assessment, technical transparency alone is not sufficient; the reasoning must follow the structure of the legal evaluation it supports \citep{herrmann2026siak,herrmann2026embedding}. Black-box decision-making may also raise concerns under the fair trial guarantees of Article 6 ECHR and Article 47 of the Charter of Fundamental Rights \citep{echr_art6,cfreu_art47}, and the EU AI Act emphasizes transparency for high-risk AI systems used in law enforcement \citep{euaiact2024}.

\textbf{R5.} Interoperability and Extensibility: LEAs run heterogeneous technical environments. The system must work with existing forensic toolchains and data formats and must accommodate changing investigative needs without architectural change. A relevant constraint in this setting is that forensic IT environments are typically separated, physically or logically, from operational networks. Federated learning therefore requires controlled mechanisms for transferring model updates across these boundaries without compromising the isolation of the forensic environment.

\textbf{R6.} Integration with Forensic Workflows:
FL should extend established workflows rather than disrupt them. The system must fit evidence management, case handling, and analytical pipelines, and must allow incremental adoption where that is useful. Integration must also preserve the distinction between intelligence and evidence. A model output may direct the collection of evidence following the forensic procedures described in ISO/IEC 27037, but it does not replace that collection.

\textbf{R7.} Usability and Operational Practicality:
The system must be usable by non-specialists within normal agency resource constraints. This includes deployment: installation, configuration, and maintenance must be feasible for an agency IT department without specialist machine learning support, because a system that cannot be set up will not be adopted, whatever its properties. Excessive complexity is also more than an adoption barrier in this setting. It can make the system harder to operate reliably and its behavior more difficult to explain or review. This is particularly relevant where model outputs influence investigative decisions and may later require scrutiny.

\begin{table}[th]
\caption{Additional demands placed on each requirement by the forensic setting, relative to its formulation in federated learning for healthcare and finance.}
\label{tab:domain_contrast}
\centering
\scriptsize 
\begin{tabular}{@{}p{0.03\linewidth}p{0.35\linewidth}p{0.52\linewidth}@{}}
\toprule
\textbf{Req.} & \textbf{Formulation in regulated domain FL} & \textbf{Additional demand in the forensic setting} \\
\midrule

R1 & Raw data remains local; secure aggregation and differential privacy protect individual records against an aggregator with no claim to them. & Purpose limitation may apply at the level of the individual investigation; the protection required may therefore differ by layer depending on the applicable legal and operational requirements \citep{gdpr_article5,directive2016680}. \\
\addlinespace
R2 & Logging for reproducibility, model governance and regulatory reporting. & The record must remain verifiable over time and withstand independent or critical examination (ISO/IEC 27041 and 27043). \\
\addlinespace
R3 & Consortium agreements between voluntarily cooperating institutions. & Participation may be constrained by statutory competence rather than determined solely by voluntary agreement. \\
\addlinespace
R4 & Explanations addressed to domain experts and regulators. & Explainability requirements may extend beyond the system operator where model outputs influence investigative or judicial processes; the required level of court-facing explainability remains context-dependent \citep{echr_art6,cfreu_art47}. \\
\addlinespace
R5 & Integration with existing institutional data infrastructure. & Integration across networks that are deliberately separated from operational infrastructure for security reasons. \\
\addlinespace
R6 & Integration with clinical or business processes. & Preservation of the boundary between intelligence and evidence; outputs may direct evidence collection following established forensic procedures rather than replacing it (ISO/IEC 27037). \\
\addlinespace
R7 & Usability for domain specialists. & System complexity must remain manageable for reliable operation, review, and explanation, particularly where outputs influence investigative decisions. \\

\bottomrule
\end{tabular}
\end{table}

\subsection{Tensions between requirements}
\label{subsec:req_tensions}

The requirements cannot be met independently of one another; three tensions arise:

One conflict is between R1 and R2. Differential privacy adds calibrated noise to model updates so that individual contributions cannot be inferred, and secure aggregation hides individual updates from the aggregator. Both degrade exactly the property that R2 needs, namely the ability to attribute a contribution to a specific participant afterwards. In healthcare and finance, this tension stays latent, because no external party has an interest in contesting the provenance of one institution's contribution.

The conflict is narrower than it looks because R1 is interpreted differently at each layer. It only appears where a privacy mechanism is needed, and it may become particularly relevant where the aggregate leaves the jurisdiction. Inside a jurisdiction the aggregating body may itself be a competent authority, but the protection required still depends on the purpose and legal basis of the processing and on what information is necessary for that purpose. Privacy-preserving mechanisms can minimize the information disclosed, but they do not replace the legal basis required for the underlying processing. The open question is therefore not how to reconcile privacy and auditability in general. It is whether attribution can be kept across those aggregation steps where additional privacy protection is needed. We raise the question here and do not answer it. Designing an audit mechanism that satisfies both requirements is a problem of its own and is treated in separate work.

Two further tensions are less severe but relevant for deployment. R4 partly opposes R1, because explanations of model behavior can themselves disclose information about the training data. R2 also conflicts with R7, as comprehensive audit trails burden agencies already short of resources.

\subsection{Federated learning framework landscape: Screening and forensic readiness}
\label{subsec:Federated Learning Framework Landscape: Maturity and Forensic Readiness}

The architecture in Sec.~\ref{Federated Forensics: A Reference Architecture for Law Enforcement Collaboration} needs hierarchical aggregation and a point to inspect contributions before forwarding. Most maintained FL frameworks offer the first, and the second can be built wherever aggregation is user-definable, so the architecture is not tied to one framework.

For completeness, we screened twelve widely used open-source frameworks on maintenance indicators taken from public repository metadata, and then assessed one of them, Flower, against R1 to R7. The screening procedure, the statistics, the reasons for choosing Flower, and the assessment itself are in Appendix~\ref{appendix:maturity} and Appendix~\ref{appendix:flower}. Its indicators measure public community activity, a poor proxy when adopting organizations do not contribute publicly, and they also depend on project age and the developer community's location.

One finding from that assessment belongs here, because it bears on the argument and not on the framework. Flower provides secure aggregation and differential privacy, but both are configured globally rather than per aggregation step, which reflects the flat deployment model the framework assumes. As argued under R1, the question for a forensic deployment is which protection is required at each layer based on the applicable legal and operational requirements. A framework with a single privacy setting for all aggregation steps cannot express it, and enabling either mechanism everywhere may trigger the conflict with auditability described in Sec.~\ref{subsec:req_tensions} at layers where the same level of protection is not required.

The remaining findings agree: several frameworks are mature, but none offers inter-agency governance, forensic-grade audit trails, or case-management integration. Standard FL frameworks were not designed for the hierarchies, oversight, and accountability demands of law enforcement, and this mismatch motivates the architecture in the next section.

\section{Federated Forensics: A reference architecture for law enforcement collaboration}
\label{Federated Forensics: A Reference Architecture for Law Enforcement Collaboration}

\emph{Federated Forensics} works with existing FL technology and mirrors the structure of law enforcement organizations \citep{buerger2023}, because that structure defines responsibility boundaries. It is deliberately minimal: it fixes where coordination and enforcement happen, not how each requirement is implemented, and specifies no mechanisms for auditability or explainability.

\subsection{Three-layer Federated Forensics architecture}

\begin{figure}[th]
    \centering    
    \includegraphics[width=1.0\linewidth]{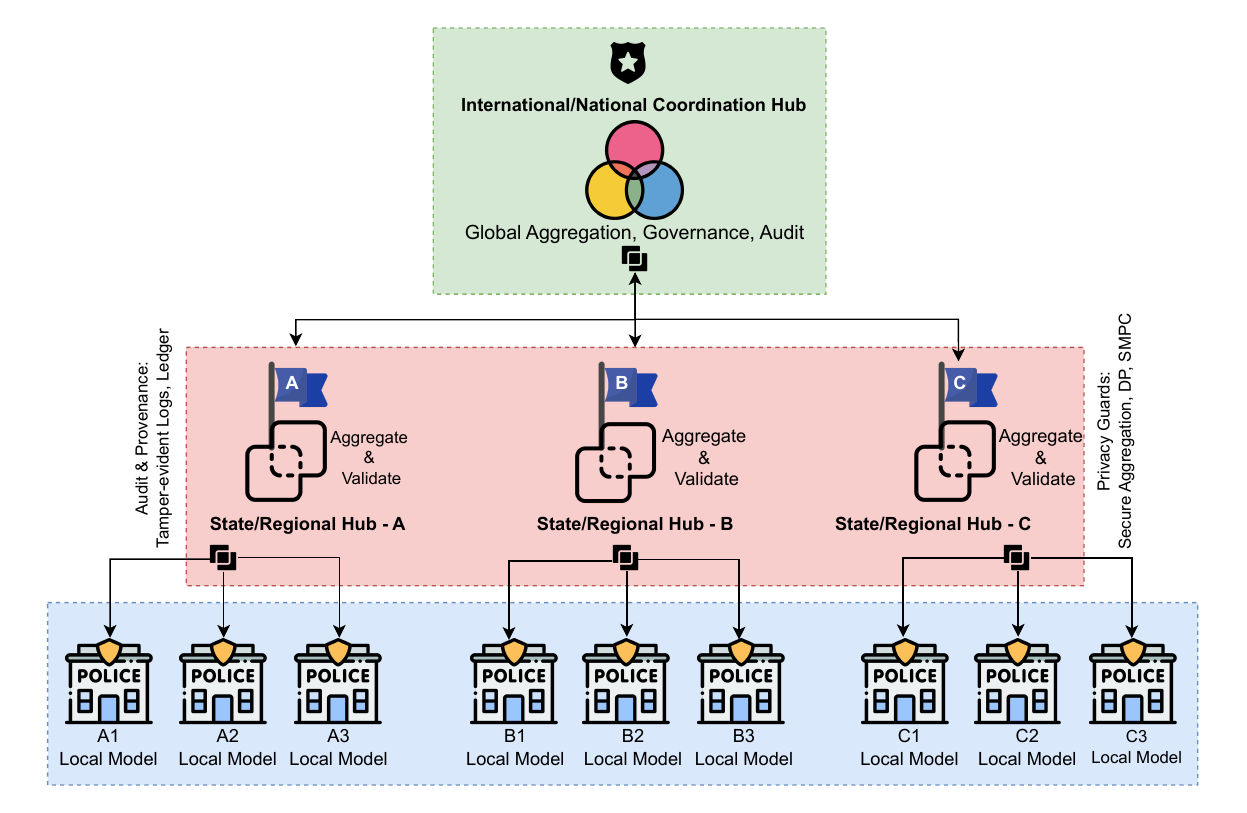}
 
   \caption{Conceptual architecture for \emph{Federated Forensics}. Local nodes train models and send protected updates to regional hubs, which validate, aggregate, and forward to the national coordinator with privacy guards and audit mechanisms.}
    
    \label{fig:fedforensic_arch}
\end{figure}

We propose three logical layers, shown in Fig.~\ref{fig:fedforensic_arch}.
The \emph{Local Agency Layer} contains the individual LEAs that collect, store, and analyze evidence. All raw data stays here. Agencies train models on their own evidence and share only model updates upwards, so FL extends existing forensic workflows rather than replacing local analysis.

The \emph{State/Regional Coordination Layer} corresponds to intermediate entities such as state police units, inter-agency task forces, or regional forensic hubs. It receives updates from local agencies, aggregates them, and produces two outputs: the \emph{regional model}, which may draw on every contribution the layer is permitted to use, and the \emph{outward contribution}, which carries only what may cross the jurisdictional boundary. The outward contribution is not a usable model of its own. It is an intermediate aggregate, passed on for further aggregation at the national layer. The component that separates the two is the \emph{policy gate}. It reads the policy label of an update and decides, per round and per recipient, whether the update may enter the aggregate. Sec.~\ref{subsec:policy_enforcement} explains why this decision belongs at this layer and where the labels come from.

Models return the same way in reverse: national to regional to local. A region that forwarded everything passes the national model down unchanged. A region that withheld something first combines the national model with its regional model, so its agencies still benefit from the withheld knowledge without it leaving the region (Sec.~\ref{Prototype Implementation} gives the combination rule). Each agency then trains its next round on the received model, and the resulting updates are labeled and gated again.

The \emph{National/International Coordination Layer} aggregates the regional contributions for strategic intelligence and cross-border pattern detection, under existing national or international governance. It sees only what a region forwarded, not what it withheld. Reporting exclusions upwards would reveal that a region runs operations of a restricted category, the very inference Sec.~\ref{subsec:threat_model} guards against, so exclusions are recorded only at the regional layer.

\subsection{Trust assumptions and threat model}
\label{subsec:threat_model}

Local agencies are \emph{semi-trusted}: they follow the protocol under normal circumstances, but they are subject to insider threat, credential compromise, and coercion. The regional and national layers are \emph{honest-but-curious}: they aggregate correctly, but they may try to derive information from what they see, and they are high-value targets. The communication between layers is authenticated and encrypted. Consequently, Table~\ref{tab:threat_model} states what each party observes, what it can infer beyond that, and what it can influence. The inference column matters most, because it describes what an agency discloses by taking part at all.

Three points deserve attention. First, metadata stays informative even when update contents are protected. Participation timing and declared sample counts let a regional hub infer the relative caseload of its members, and a long absence can itself indicate a sensitive ongoing operation. Encrypting the updates does not help against this.

Second, the protection owed depends on the layer. A local agency and its regional hub belong to the same jurisdiction and often stand in a relationship of subordination. The legal basis and the protection required still depend on the specific processing purpose. What matters here additionally is operational secrecy: an agency may accept that the hub processes its contribution, but not that a neighboring agency learns which operations it runs. Hiding contributions from peers serves this; adding noise does not.

At the next layer, the recipient may be a foreign or supranational body, so the legal basis for any transfer must be assessed separately, and data minimization may demand extra protection. If noise is justified anywhere, it is here. The structure already helps, since this layer receives only multi-agency aggregates, never an individual site's update. One risk remains at every layer: a model that has memorized rare training examples can leak them when redistributed, which matters for sensitive investigative material.

Third, the last row of Table~\ref{tab:threat_model} has no counterpart in healthcare or finance. Forensic training data comes largely from the subjects of the investigation: seized devices, intercepted communications, and material published by criminal groups themselves. These subjects are adversaries who may change their behavior if they suspect they are being modeled. A patient does not alter a radiograph to mislead a diagnostic model; a criminal group has both the motive and the means to alter its communications. This threat originates outside the system boundary, and no mechanism inside the system can address it.

\begin{table*}[htpb]
\caption{What each party observes, can infer, and can influence in the three-layer architecture.}
\label{tab:threat_model}
\centering
% \footnotesize
\scriptsize
\begin{tabular}{@{}p{0.13\textwidth}p{0.23\textwidth}p{0.29\textwidth}p{0.25\textwidth}@{}}
\toprule
\textbf{Party} & \textbf{Observes directly} & \textbf{Can infer} & \textbf{Can influence} \\
\midrule
Local agency & Own evidence and local model; global model received & Aggregate properties of other regions, from successive global models & Own update; own declared sample count \\
\addlinespace
Regional hub & Member identities; per-agency updates; sample counts; participation timing & Relative caseload per agency; properties of local training data from gradients; sensitive operations, from participation gaps & Regional aggregate; inclusion, exclusion and weighting of members \\
\addlinespace
National hub & Regional aggregates; region identity; aggregate weights & Relative data volume per region, but not individual contributions & Global model redistributed to all participants \\
\addlinespace
Network adversary & Encrypted traffic; timing and volume & Participation patterns; periods of regional activity & Availability \\
\addlinespace
Insider at agency & Local raw evidence; credentials & Contents of the global model & Poisoned updates; exfiltration of the global model \\
\addlinespace
Investigated party & Nothing within the system & None & The training data itself, by adapting behavior if monitoring is suspected \\
\bottomrule
\end{tabular}
\end{table*}

\subsection{Jurisdictional policy enforcement}
\label{subsec:policy_enforcement}

Consider a restriction found in practice. Under the national implementation of Directive (EU) 2016/680, a German agency may hold cases whose material must not contribute to a model received by other jurisdictions \citep{directive2016680}. A flat topology can comply only by leaving that material out of training. Since a flat deployment has one shared model, the material then also disappears from every model used \emph{inside} the jurisdiction, where its use may be lawful and neighboring agencies work on related cases. Compliance thus discards more than the restriction requires. The withholding is also invisible: it happens inside each agency, so no agency can show the others that it complied.

The three-layer structure separates what an agency trains on from how far the resulting signal travels: restricted material improves the regional model, and the policy gate withholds it from the global one (Fig.~\ref{fig:gate_comparison}).

The gate needs labels with a clear origin and granularity. One workable choice is the case category, the level at which legal rules are often written and case management systems already classify holdings, so labels come from the case file rather than an ad hoc decision at training time. Other levels, such as the investigation or data source, work the same way; our prototype uses a two-valued label per data item (Sec.~\ref{Prototype Implementation}). Each update carries exactly one scope, so an agency submits one update per scope and round. This costs local training effort and shrinks each training set, but a mixed update would have to be treated under its most restrictive label, losing the utility the architecture exists to preserve.

Two limits apply. Mapping legal rules to case categories is a legal decision made once per jurisdiction; the architecture makes it explicit and reviewable but does not make it. And the gate depends on its inputs: a mislabeled case is processed under the wrong policy, and the regional layer cannot detect this. The gate's decisions are inspectable; the labels' correctness is not guaranteed.

The more important argument concerns authority. A lawful-transfer decision must be taken by a competent body. A flat topology offers two enforcement points, and both fail: the agency is competent but cannot scope or demonstrate compliance, and the global server could enforce visibly but may be controlled by another jurisdiction or an international body, which lacks the authority to decide. The regional layer is the only point that is both competent and a single place where each decision is recorded. This alignment, not the aggregation algorithm, is what the architecture contributes. Its cost, reduced information for the global model, is quantified in Sec.~\ref{subsec:empirical_results}.

\subsection{Architectural implications for forensic requirements}
\label{subsec:arch_implications}

The seven requirements are not properties of the same kind, so they cannot all be assessed against an architecture in the same way. Two of them are settled by the structure alone. Raw data cannot leave the local layer because no higher layer receives it (R1), and the authority over each aggregation step is fixed by the layer that performs it, which is where the policy gate acts (R3). For two more, the structure says where, but not how. The aggregation points fix where audit records must be produced, but say nothing about what such a record should contain or how it is verified (R2), and the boundary between intelligence and evidence sits at the local layer, but integration with evidence management systems is not demonstrated (R6). For the remaining three, the structure makes no difference: explainability follows from the choice of model (R4), interoperability from the toolchain and deployment decisions (R5), and usability from implementation, interface design, and training (R7).
No topology can deliver usability or explainability, and claiming otherwise would overstate what an architecture contributes.

\subsection{Alternative architectural approaches}
\label{Alternative Architectural Approaches}

We judge other topologies by one test: does an aggregation point sit at a legal boundary, and can it hold updates back?

\emph{Flat cross-silo} architectures fail the criterion twice, as shown in Sec.~\ref{subsec:policy_enforcement}: their one aggregation point lies outside every participating jurisdiction's competence, and agency-side withholding cannot scope.

\emph{Parallel flat federations} obtain the scoping effect without a hierarchy: an agency joins one federation inside its jurisdiction and a second one across borders, and contributes restricted material only to the first. Where two federations are enough, this is a legitimate alternative, with three drawbacks. Training and coordination run twice, and the cost grows with the number of scopes. The two federations never meet, so the inner model gains nothing from what the cross-border model learns. And the decision of which federation a contribution enters is still taken inside each agency, so the enforcement point stays away from the competent authority and outside any single point of record. Parallel federations rebuild the hierarchy by hand, without its enforcement point.

\emph{Two-layer} architectures, in which agencies report directly to a national or international coordinator, are sufficient where the relevant boundary of competence is national. Where oversight sits at state or regional level, as in federal police structures \citep{buerger2023}, the enforcement point and the competent authority no longer coincide.

\emph{Fully decentralized} approaches have no distinguished aggregation point at all, so there is no place at which a restriction can be applied and no party responsible for applying it.

The three-layer structure is therefore not optimal in general. It is the minimal structure that aligns an enforcement point with legal competence in the federal and cross-border settings we consider. Where the competence structure has more levels, further layers follow by the same argument, though we have not examined their cost.
\section{Prototype implementation and empirical results}
\label{Prototype Implementation}

This section tests, in a controlled proof-of-concept experiment, whether the policy gate keeps restricted updates inside their jurisdiction, and at what cost. Malware classification is only the vehicle; we propose no new classifier. The codebase is available in the repository.
\footnote{\url{https://github.com/ArunaSanjeewa/federated-forensics}}
ßß
\subsection{Setup}
\label{sec:setup}

\paragraph{Prototype layout}

% Figure for Sec. 4: update flow, policy gate, and return path, flat vs. gated.
% Preamble needs:  \usepackage{tikz}  \usetikzlibrary{positioning,calc}
\begin{figure}[t]
% \centering
\scriptsize
\begin{tikzpicture}[
  scale=0.65,
  transform shape,
  ag/.style={draw, rounded corners=1pt, minimum width=11mm, minimum height=6mm, font=\scriptsize},
  hub/.style={draw, thick, rounded corners=1pt, minimum width=26mm, minimum height=7mm, font=\scriptsize, align=center},
  chip/.style={draw, minimum width=4.2mm, minimum height=3.4mm, font=\tiny, inner sep=1pt},
  chipr/.style={chip, fill=gray!35},
  lbl/.style={font=\scriptsize\itshape, align=center, text width=64mm},
  elbl/.style={font=\tiny},
  up/.style={->, >=stealth},
  down/.style={->, >=stealth, densely dashed, gray},
  bnd/.style={densely dotted, thick}
]
%================ (a) FLAT ================
\begin{scope}
\node[ag] (fa1) at (0,0) {A1};
\node[ag] (fa2) at (2.0,0) {A2};
\node[ag] (fa3) at (4.0,0) {A3};
% data chips
\node[chip, below=1mm of fa1.south, xshift=-2.6mm] (fg1) {g};
\node[chipr, right=0.8mm of fg1] (fr1) {r};
\draw[red!55!black, semithick] (fr1.south west) -- (fr1.north east);
\node[chip, below=1mm of fa2.south] {g};
\node[chip, below=1mm of fa3.south] {g};
% global server above the boundary
\node[hub] (fg) at (2.0,2.9) {Global server};
\draw[bnd] (-1.2,1.7) -- (5.4,1.7);
\node[elbl, gray, anchor=east, align=right] at (-0.25,2.0) {jurisdictional\\boundary};
% updates
\draw[up] (fa1.north) -- node[elbl, left=1pt, pos=0.35] {$U_g$} (fg.south -| fa1.north);
\draw[up] (fa2.north) -- node[elbl, left=1pt, pos=0.35] {$U_g$} (fg.south);
\draw[up] (fa3.north) -- node[elbl, left=1pt, pos=0.35] {$U_g$} (fg.south -| fa3.north);
% return
\draw[down] (fg.east) .. controls (5.6,2.6) and (5.6,1.0) .. (fa3.east);
\node[elbl, gray, anchor=east, align=right] at (5.25,1.25) {global\\model};
\node[lbl] at (2.0,-1.55) {(a) Flat federation: A1's restricted data (r) is excluded from training and improves no model.};
\end{scope}
%================ (b) GATED ================
\begin{scope}[xshift=75mm]
\node[ag] (ha1) at (0,0) {A1};
\node[ag] (ha2) at (2.8,0) {A2};
\node[chip, below=1mm of ha1.south, xshift=-2.6mm] (hg1) {g};
\node[chipr, right=0.8mm of hg1] {r};
\node[chip, below=1mm of ha2.south] {g};
% hub, boundary, national
\node[hub] (hr) at (1.4,2.0) {Regional hub\\[-1pt]\tiny policy gate $\cdot$ regional model $(g{+}r)$};
\draw[bnd] (-1.2,3.15) -- (4.5,3.15);
\node[elbl, gray, anchor=east, align=right] at (-0.25,3.45) {jurisdictional\\boundary};
\node[hub] (hn) at (1.4,4.35) {National server};
% updates up
\draw[up] (ha1.north) -- node[elbl, left=1pt, pos=0.4] {$U_g,\,U_r$} (hr.south -| ha1.north);
\draw[up] (ha2.north) -- node[elbl, right=1pt, pos=0.4] {$U_g$} (hr.south -| ha2.north);
% outward contribution crosses the line; r stops at it
\draw[up] ([xshift=5mm]hr.north) -- node[elbl, right=1pt, pos=0.62, align=left] {outward\\contribution $(g)$} ([xshift=5mm]hn.south);
\draw[up, -] ([xshift=-7mm]hr.north) -- node[elbl, left=1pt] {$U_r$} ([xshift=-7mm]hr.north |- 0,3.15);
\draw[thick] ([xshift=-8.6mm]hr.north |- 0,3.15) -- ([xshift=-5.4mm]hr.north |- 0,3.15);
% return path
% \draw[down] (hn.east) .. controls (5.0,4.0) and (5.0,2.6) .. (hr.east);
\draw[down] (hn.east) .. controls (4.4,4.0) and (4.4,2.6) .. (hr.east);
\node[elbl, gray, anchor=west, align=left] at (3.9,4.0) {global\\model};
\draw[down] ([xshift=-8mm]hr.south) -- ([xshift=-8mm]hr.south |- ha1.north);
\draw[down] ([xshift=10mm]hr.south) -- ([xshift=10mm]hr.south |- ha2.north);
\node[elbl, gray, align=center] at (1.4,0.88) {combined\\model};
\node[lbl] at (1.4,-1.55) {(b) Policy-gated hierarchy: r improves the regional model, the gate stops $U_r$ at the boundary, and agencies start the next round from the combined model.};
\end{scope}
\end{tikzpicture}
\caption{Update ($U$) flow and return path for a restricted contribution. Agency A1 holds material of a global scope ($g$) and of a restricted scope ($r$) and trains one update per scope. In a flat federation (a), compliance forces A1 to exclude $r$ from training entirely. In the gated hierarchy (b), $U_r$ enters the regional model, the policy gate withholds it from the outward contribution at the jurisdictional boundary, and the returned model, national combined with regional, is the starting point of the next round.}
\label{fig:gate_comparison}
\end{figure}

The system runs as a single-process simulation with four local
agency nodes (A1 to A4), grouped into two regions
(R1 $=$ \{A1, A2\}, R2 $=$ \{A3, A4\}), reporting to one
national node. 
Each node keeps its own model; the regional and national nodes aggregate, and the regional nodes also apply the policy gate. In the proposed
condition the two agencies holding restricted data (A1, A2) each send two separate update
streams per round, while A3 and A4 send one each;
so the regional node in R1 receives four update streams per round and the
regional node in R2 receives two, six in total across the federation.

\paragraph{Dataset}
We use BODMAS (Blue Hexagon Open Dataset for Malware Analysis;
\citet{yang2021bodmas}), a public dataset for malware research. It contains 134'435
rows (57'293 malware, 77'142 benign). 
We use only the malware rows and classify each sample's feature vector into its family. Of the 582 families, we use eight, chosen because each has enough samples to fill every entitled agency's 400-row budget and each \texttt{REGIONAL\_ONLY} family has a \texttt{GLOBAL} sibling in the same category (Table~\ref{tab:families}; Sec.~\ref{sec:policy}). 
The dataset only serves to demonstrate the gate. Realistic digital forensic datasets are rarely available \citep{breitinger2023sharing}, since evidentiary material is legally protected and seldom shared, whereas BODMAS is public, large, and labeled. The results therefore make no claim about any real investigation, but show that the architecture enforces a jurisdictional restriction as designed and at a measurable cost.

\begin{table}[t]
\centering
\scriptsize
\caption{Eight classes; \texttt{trojan} and \texttt{worm} each contain both a
\texttt{GLOBAL} and a \texttt{REGIONAL\_ONLY} family, so restricting
gepys/sillyp2p removes two specific families, not the trojan/worm
categories. The mapping is a design choice made before any run.}
\label{tab:families}
\begin{tabular}{llr}
\toprule
Propagation scope & Family (category) & Samples in BODMAS \\
\midrule
\texttt{GLOBAL} & wacatac (trojan)   & 4'694 \\
\texttt{GLOBAL} & dinwod (trojan)    & 2'057 \\
\texttt{GLOBAL} & sfone (worm)       & 4'729 \\
\texttt{GLOBAL} & mira (worm)        & 1'960 \\
\texttt{GLOBAL} & wabot (backdoor)   & 3'673 \\
\texttt{GLOBAL} & berbew (backdoor)  & 1'749 \\
\texttt{REGIONAL\_ONLY} & gepys (trojan)    & 1'124 \\
\texttt{REGIONAL\_ONLY} & sillyp2p (worm)   & 1'616 \\
\bottomrule
\end{tabular}
\end{table}

\paragraph{Splitting into agencies and regions}
The data is distributed in three steps. 
First, each family receives a propagation scope (Table~\ref{tab:families}): \texttt{GLOBAL} updates may reach the national level; \texttt{REGIONAL\_ONLY} updates may not leave their region. 
Second, each family's rows are shuffled and cut into disjoint blocks of 400 rows. Every agency receives one block of each \texttt{GLOBAL} family, but only A1 and A2 receive a block of each \texttt{REGIONAL\_ONLY} family. A1 and A2 thus hold eight families (3'200 rows); A3 and A4 hold six (2'400 rows). This assignment is fixed by one partition seed, so every agency owns the same rows in every run. 
Third, each agency splits every block into 300 training and 100 test rows. This split is redrawn in each of the five repetitions (seeds 42 to 46), so A1 and A2 always have restricted rows in both training and test sets.

\paragraph{Model and training}
Every condition in Sec.~\ref{sec:conditions} trains the same machine learning model from scratch: a multi-layer perceptron, a simple neural network, with layer widths 2'381, 256, 64, and 8, matching the input features and the eight families. It has 626'760 parameters, uses ReLU activations and dropout of 0.1, and is trained in 32-bit floating point. The whole network is trained and exchanged every round. Features are $z$-score normalized with statistics computed over each seed's pooled training rows (Sec.~\ref{sec:limitations}). Training uses the AdamW optimizer with learning rate $1\times10^{-3}$, batch size 32, three local epochs per round, and gradient-norm clipping at 1.0. Federation runs for five rounds with sample-weighted FedAvg, identical for B2 to B4 except for one step in B4 (Sec.~\ref{sec:policy}). Each condition runs with five seeds (42 to 46).

\paragraph{Technical stack used}
The federated learning framework is Flower (\texttt{flwr}~1.36.0), with
PyTorch~2.14.0 and scikit-learn~1.9.1 on Python~3.13.5; paired
$t$-tests use SciPy~1.18.1. Runs use an Apple M4~Pro machine (48\,GB
memory).

\subsection{Policy configuration}
\label{sec:policy}

\paragraph{What ``restricted'' means here}
Two of the eight malware families are \texttt{REGIONAL\_ONLY}, i.e., updates trained on them may reach regional partners but not the national level. 
Each has a \texttt{GLOBAL} sibling in its category (gepys with the trojan wacatac, sillyp2p with the worm sfone), so excluding them does not remove the trojan or worm categories from a compliant model.

\paragraph{Who holds it}
Only the two Region~1 agencies, A1 and A2, hold restricted material, 25\% of their data (800 of 3'200 rows).

\paragraph{How the label is attached to an update}
Each agency declares the scope of its data on every update it sends, together with an update id, agency, region, round, number of training examples, and a parameter hash. A1 and A2 send two updates per round, $U_g$ from their \texttt{GLOBAL} rows and $U_r$ from their \texttt{REGIONAL\_ONLY} rows. The regional hub applies the gate to the declared scope, excluding only \texttt{REGIONAL\_ONLY} updates bound for the national level, and records every decision, \texttt{INCLUDE} or \texttt{EXCLUDE}, in an append-only provenance log, so blocked updates are documented rather than silently dropped.

\paragraph{Return path}
Region~2 agencies receive the national model unchanged. Region~1 agencies receive an equal (50/50) blend of the national model and their region’s aggregate of the $U_r$ updates. We do not weight the blend by example count, because the national model never sees the two restricted families and therefore learns to rule them out. Only the $U_r$ aggregate knows them, and it is based on far fewer examples than the national model. Weighted by example count, it would be outvoted and the restricted families largely forgotten; equal weighting keeps them.

\subsection{The four experiment run conditions}
\label{sec:conditions}

\begin{description}
\item[B1 - Local only.] Each agency trains on its own data and never shares.

\item[B2 - Flat FL, unrestricted.] 
One flat federation over all data, including restricted data, shared everywhere. Not legally available in practice: it puts restricted knowledge into the national model.

\item[B3 - Flat FL, restricted dropped.] One flat federation, but A1 and A2
must leave their \texttt{REGIONAL\_ONLY} data out of training entirely. This is
what an agency can do today if it wants to join a flat federation and stay
compliant.

\item[B4 - Hierarchical, policy aware (our proposal).] Three layers.
Restricted data contributes to the Region~1 model through the separate
$U_r$ stream; the gate blocks that stream at the
region-to-national boundary; the return path keeps it inside Region~1.
\end{description}

Two comparisons matter, both measured at A1 and A2, the agencies holding restricted material: B4 against B3 shows what our architecture gains over the only compliant option today, and B2 against B4 shows what complying with the restriction costs. In addition, A3 and A4 show whether restricted knowledge leaks to agencies that never held it, and B1, where each agency trains alone, serves as a reference for whether collaboration pays off at all.

\subsection{Evaluation protocol}
\label{sec:protocol}

We report results per agency rather than as one overall figure, since an investigator cares about what a model does for a specific agency, and the restricted material sits in only one region. Each agency's model is tested on two sets: the agency's own held-out samples (600 to 800), and the restricted samples of A1 and A2 combined (400). We apply the second set to A3 and A4 as well, to see whether restricted capability leaks to agencies that never held the material.

The main metric is macro-F1, which averages the F1 score over the malware families present in a test set, so that a rare family weighs as much as a common one. On the restricted set, only the two restricted families are present, and only these are averaged. We also report accuracy.

For B4, we count every restricted update that reaches the regional node, and check against the provenance log whether the gate blocked it or let it through. Each condition is run five times with a new train/test split, and all runs use the same scoring code. We report the mean and standard deviation (SD) over the five runs and test differences between conditions with paired $t$-tests ($p$).
\\

\subsection{Empirical results}
\label{subsec:empirical_results}

\paragraph{Policy enforcement}

Across all five runs, the gate blocked each of the 50 restricted updates bound for the national level and passed all 100 \texttt{GLOBAL} updates, with a matching \texttt{EXCLUDE} record in the provenance log for every block. The implementation enforces the rule as specified, so the following results come from a system that actually complies.

\paragraph{Classification performance}

Table~\ref{tab:peragency} shows macro-F1 per agency and condition, on each agency's own test set and on the restricted test set; Table~\ref{tab:headline} gives the main comparisons.

\begin{table}[t]
\centering
\scriptsize
\caption{Macro-F1 per agency (mean $\pm$ SD, five runs), on each agency's own test set and on the restricted test set of A1 and A2. A3 and A4 hold no restricted data; their restricted-set scores show leakage.}
\label{tab:peragency}
\begin{tabular}{lcccc}
\toprule
Agency & B1 & B2 & B3 & B4 \\
\midrule
\multicolumn{5}{l}{\emph{Own test set}} \\
A1 & 0.919 $\pm$ 0.009 & 0.930 $\pm$ 0.009 & 0.632 $\pm$ 0.012 & 0.913 $\pm$ 0.009 \\
A2 & 0.923 $\pm$ 0.005 & 0.917 $\pm$ 0.008 & 0.626 $\pm$ 0.007 & 0.891 $\pm$ 0.006 \\
A3 & 0.891 $\pm$ 0.018 & 0.896 $\pm$ 0.019 & 0.899 $\pm$ 0.021 & 0.896 $\pm$ 0.020 \\
A4 & 0.905 $\pm$ 0.020 & 0.905 $\pm$ 0.016 & 0.905 $\pm$ 0.015 & 0.907 $\pm$ 0.016 \\
\midrule
\multicolumn{5}{l}{\emph{Restricted test set}} \\
A1 & 0.991 $\pm$ 0.006 & 0.986 $\pm$ 0.006 & 0.000 & 0.962 $\pm$ 0.014 \\
A2 & 0.994 $\pm$ 0.004 & 0.986 $\pm$ 0.006 & 0.000 & 0.962 $\pm$ 0.014 \\
A3 & 0.000 & 0.986 $\pm$ 0.006 & 0.000 & 0.000 \\
A4 & 0.000 & 0.986 $\pm$ 0.006 & 0.000 & 0.000 \\
\bottomrule
\end{tabular}
\end{table}

\begin{table}[t]
\centering
\scriptsize
\caption{Comparisons on the restricted test set (A1, A2; identical
by construction, since they share the Region~1 model and the pooled test
set; see Sec.~\ref{sec:limitations}). Five runs, paired $t$-test with four
degrees of freedom. Macro precision is exactly 1.000 for both B2 and B4 in
every run.}
\label{tab:headline}
\begin{tabular}{llrr}
\toprule
Comparison & Metric & Difference & $p$ \\
\midrule
B4 $-$ B3 & accuracy   & $+0.930 \pm 0.025$  & $<0.0001$ \\
(main comparison) & macro-F1   & $+0.962 \pm 0.014$  & $<0.0001$ \\
                  
\midrule
B2 $-$ B4 & accuracy   & $+0.044 \pm 0.020$  & $0.008$ \\
(cost of compliance) & macro-F1 & $+0.024 \pm 0.012$  & $0.010$ \\
                 
\bottomrule
\end{tabular}
\end{table}

\paragraph{Main result (B4 against B3)}
B3 cannot recognize the restricted families at all: it scores 0\% accuracy on them in every run, because they were never part of its training. Since a quarter of A1's and A2's caseload belongs to these families, their macro-F1 on their own test sets drops from about 0.91 to 0.63. B4 restores this capability, reaching 93.0\% accuracy and 0.962 macro-F1 on the restricted families ($p<0.0001$). Its precision there is 1.000 in every run: it never confuses gepys with sillyp2p, and its errors are misclassifications into one of the six non-restricted families.

In practical terms, both options obey the restriction, but they leave the agency with very different tools. In a flat federation, the agency's model cannot recognize the very cases the restriction concerns; with the gate, it recognizes them in 93\% of cases.

\paragraph{Cost of compliance (B2 against B4)}
B2, which ignores the restriction, outperforms B4 on the restricted families by 4.4 percentage points in accuracy ($p=0.008$) and 0.024 in macro-F1 ($p=0.010$). On the agencies' own test sets, the cost is 0.017 macro-F1 for A1 ($p=0.0008$) and 0.026 for A2 ($p=0.0002$), and there is no measurable cost for A3 and A4.

In practical terms, respecting the restriction means that out of 100 samples from the restricted families, the agency's model misclassifies about four more than a model that ignores the rule would. Agencies without restricted material see no difference.

\paragraph{Leakage}
B2's advantage comes at a price that the restriction exists to prevent. Because all agencies share one global model, A3 and A4, which hold no restricted material, recognize the restricted families as well as A1 and A2 do (0.986 macro-F1). Under B4, they score 0.000, so the capability stays inside Region~1.

In practical terms, ignoring the rule lets agencies that never held the restricted material recognize it anyway. With the gate, only the agencies entitled to that knowledge gain it.

\paragraph{Comparison with local training (B1)}
An agency training alone recognizes its own restricted families slightly better than under B4 (0.991 and 0.994 against 0.962; $p<0.01$), since it uses all its data and blends in no outside model. On their own test sets, B4 is on par with B1 for A1 (difference 0.006, not significant) and 0.032 below it for A2 ($p=0.0016$). But B1 gives up all collaboration: no agency benefits from another, and A3 and A4 cannot recognize anything they do not hold themselves.

In practical terms, an agency with enough data of its own may do slightly better alone on its own cases. But it then learns nothing from other agencies, and they learn nothing from it.

\paragraph{Summary}
The gate blocked every restricted update with a complete audit trail. B4 restores recognition of the restricted families from 0\% under B3 to 93.0\%, 4.4 points below B2, costs two to three macro-F1 points on the restricted agencies' own test sets, and, unlike B2, keeps the restricted capability inside its region.

\section{Limitations}
\label{sec:limitations}

Our experiment tests the policy gate under simplified conditions. We chose which two families count as restricted; no court or law made that choice. It is also expected that B3 scores zero on these families, because a model cannot recognize what it has never seen. The more meaningful results are what B3 loses on the agencies' full caseload, how close B4 comes to B2, and how B2 spreads restricted knowledge to agencies outside the region.

Some results are less independent than they look. A1 and A2 use the same regional model, so their scores on the restricted set are one measurement, not two. The model also only chooses among eight malware families and can never answer ``benign'' or ``unknown''. This makes all scores higher than they would be in practice, so the comparisons between conditions matter, not the scores themselves. Finally, we normalized the features using the training data of all agencies together, which a real federation would not permit.

The prototype runs in a single process on one machine. All agencies behave honestly, label their data correctly, and no encryption is used. The equal blend on the return path was only tested with two regions of similar size. The experiment therefore shows that updates are routed and blocked as intended, not that a real deployment complies with the law.

Beyond the experiment, the requirements come from the literature and from  one author's work in law enforcement, not from structured interviews with investigators. The gate trusts the labels agencies declare and cannot detect wrong ones. The legal discussion stays general; a detailed analysis for specific jurisdictions needs legal experts.

\section{Conclusion}
\label{sec:conclusion}
We asked which federated learning topologies allow a jurisdictional restriction to be enforced where it legally applies, and at what cost. Our answer is that the topology matters more than the algorithm. A flat federation can only enforce such a propagation restriction by discarding restricted material completely, and it leaves the decision either with each agency, where compliance cannot be shown, or with a global server that lacks the authority to decide. Only an aggregation layer placed at the boundary of legal competence can use restricted material inside a jurisdiction and stop it at the border. In our prototype, this recovered almost all of the capability that the flat alternative loses, 93.0\% instead of 0\% accuracy on the restricted families, at a cost of 4.4 points compared to ignoring the restriction, and without spreading restricted knowledge to other regions.

This has consequences for practice. For agencies planning joint models, the first design question is not which framework or privacy mechanism to use, but where legal competence lies and whether an aggregation point coincides with it. For framework developers, the gap is concrete: 
privacy protection and forwarding decisions must be configurable per aggregation step, rather than globally across the federation. Existing frameworks provide relevant building blocks, but do not directly support this combination of layer-specific, policy-driven controls.
Where a propagation restriction applies and agencies must still collaborate across the boundary it defines, we therefore recommend \emph{Federated Forensics}, a policy-gated hierarchical architecture. It turns a complete loss into a small, measured cost; whether that cost is acceptable remains a policy decision.

Our results come from a single-process simulation on public malware data, with honest agencies and self-declared labels, and the requirements have not yet been validated with investigators. Three steps follow: structured validation of the requirements with practitioners, an audit mechanism that keeps contributions attributable where privacy protection is needed, and a deployment across separate infrastructures with real jurisdictional rules. More broadly, this work shows that legal constraints in federated learning are not something to verify after the fact, but a design input that decides which systems can be built at all.

\section*{CRediT authorship contribution statement}
\textbf{Aruna Withanage:} Conceptualization, Methodology, Software, Investigation, Writing - Original Draft, Visualization. \textbf{Jessica Herrmann:} Conceptualization, Writing - Review \& Editing. \textbf{Hudan Studiawan:} Writing - Review \& Editing. \textbf{Frank Breitinger:} Conceptualization, Writing - Review \& Editing, Supervision.
%%=====
\section*{LLM usage statement}
Some of the authors used Claude Opus 5.5 for language refinement and clarity improvement during the preparation of this manuscript. Figure~\ref{fig:gate_comparison} was also created with the assistance of Claude Opus 5.5. Claude Sonnet 5 was used for Python code improvement during prototype implementation. After using these tools, the authors reviewed and edited the content as needed and take full responsibility for the content of the publication.
%%=====
\section*{Declaration of interest}
The authors declare that they have no known competing financial interests or personal relationships that could have appeared to influence the work reported in this paper.

%%=====
\appendices

\section{Data privacy protection laws}
Table~\ref{tab:privacy_laws} lists several laws that require consideration when creating federated forensics. 

\begin{table}[ht]
\centering
\footnotesize
\caption{Data privacy protection laws of international organizations or countries (adapted from \citep{LI2023271})}
\label{tab:privacy_laws} 
\begin{tabular}{p{.9cm}p{1cm}p{5cm}}
\toprule
\textbf{Time} & \textbf{Region} & \textbf{Law}  \\
\midrule
1996 & USA & HIPAA (Health Insurance Portability and Accountability Act)  \\

2000 & Canada & PIPEDA (The Personal Information Protection and \newline Electronic Documents Act)   \\

2000 & Argentina & PDPA (Argentina's Personal Data Protection Act)   \\

2005/2017 & Japan & APPI (The Act on the Protection of Personal Information)   \\

2016 & EU & GDPR(The General Data Protection Regulation)   \\

2018 & California & CCPA (The California Consumer Privacy Act)   \\

2021 & China & PIPL (The Personal Information Protection Law)   \\
\bottomrule
\end{tabular}
\end{table}

\section{Framework screening}
\label{appendix:maturity}

Twelve open-source frameworks were screened on five indicators obtained from public repository metadata via the GitHub API\footnote{\url{https://api.github.com/repos/\{:owner\}/\{:repository\}}}: stars, contributors, creation date, latest release date, and last activity date. All values were collected on 30-09-2026 and are shown in Table~\ref{tab:fl_frameworks}.

\providecommand{\fw}[2]{%
  \begin{tabular}[t]{@{}l@{}}
    #1\\[-1pt]
    \textit{#2}
  \end{tabular}%
}

\begin{table}[t]
\centering
\caption{Open-source FL frameworks: GitHub statistics as of 30-09-2026. Stars and contributors (Contri.) are counts.}
\label{tab:fl_frameworks}
\scriptsize
\setlength{\tabcolsep}{2pt}

\resizebox{\columnwidth}{!}{%
\begin{tabular}{@{}lrrccc@{}}
\toprule
\textbf{\shortstack[l]{Framework\\(Maintainer)}} &
\textbf{Stars} &
\textbf{Contri.} &
\textbf{Created} &
\textbf{\shortstack{Latest\\release}} &
\textbf{\shortstack{Last\\activity}} \\
\midrule

\fw{Flower \citep{beutel2020flower}}{Flower}
& 7149 & 190 & 17-02-2020 & 28-09-2026 & 30-09-2026 \\[1pt]

\fw{PySyft \citep{pysyft}}{OpenMined}
& 10041 & 432 & 18-07-2017 & 13-04-2025 & 30-09-2026 \\[1pt]

\fw{FATE \citep{fate}}{WeBank}
& 6099 & 86 & 24-01-2019 & 31-07-2024 & 19-11-2024 \\[1pt]

\fw{OpenFL \citep{DBLP:journals/corr/abs-2105-06413}}{Linux Foundation}
& 843 & 109 & 12-01-2021 & 23-06-2025 & 25-08-2026 \\[1pt]

\fw{TensorFlow Federated}{Google}
& 2453 & 115 & 12-12-2018 & 26-09-2024 & 28-09-2026 \\[1pt]

\fw{FedML \citep{fedML}}{TensorOpera}
& 4067 & 66 & 21-07-2020 & 28-10-2023 & 28-10-2025 \\[1pt]

\fw{PaddleFL \citep{paddle}}{Baidu}
& 512 & 25 & 25-09-2019 & 06-12-2021 & 26-07-2023 \\[1pt]

\fw{Substra \citep{substra}}{Owkin}
& 278 & 37 & 25-10-2019 & 14-10-2024 & 14-10-2024 \\[1pt]

\fw{NVIDIA FLARE \citep{Roth_NVIDIA_FLARE_Federated_2023}}{NVIDIA}
& 975 & 78 & 23-07-2021 & 24-09-2026 & 30-09-2026 \\[1pt]

\fw{AppFL \citep{appFL,Ryu_appFLSoft}}{UChicago Argonne}
& 185 & 26 & 07-10-2021 & 25-08-2026 & 30-09-2026 \\[1pt]

\fw{SecretFlow}{Ant Group}
& 2714 & 106 & 15-04-2022 & 26-09-2025 & 24-04-2026 \\[1pt]

\fw{FederatedScope \citep{xie2022federatedscope}}{Alibaba}
& 1545 & 18 & 24-03-2022 & 03-04-2023 & 10-08-2024 \\

\bottomrule
\end{tabular}%
}
\end{table}

\section{Assessment of Flower against R1 to R7}
\label{appendix:flower}

We examine one framework in detail, to show how the requirements of Sec.~\ref{Forensic Specific Requirements for Federated Learning} translate into concrete capabilities and where they are not met. We use Flower, because its cross-silo deployment model matches the multi-agency setting and because it scored highest in a recent comparative study of FL frameworks \citep{riedel2024comparative}. Substra \citep{substra} is built for regulated multi-party collaboration and would be the conceptually closer comparator, but it is far less maintained. That trade-off between architectural fit and adoption risk is the one agencies actually face. The assessment relies on published documentation\footnote{\url{https://flower.ai/docs/framework/index.html}} rather than on a deployment, so it is indicative rather than definitive.

\textbf{R1.} Flower provides Secure Aggregation, with implementations of \texttt{SecAgg} and \texttt{SecAgg+}, and Differential Privacy. Both are configured globally rather than per aggregation step, which reflects the flat deployment model the framework assumes. This is the one finding that bears on the argument rather than on the framework, and it is discussed in Sec.~\ref{subsec:Federated Learning Framework Landscape: Maturity and Forensic Readiness}.

\textbf{R2.} Flower offers audit logging based on content-addressable messaging, which assigns cryptographic hashes to model weights and gradients by content rather than by location. The logs allow training rounds to be reconstructed technically. They are designed for system verification and debugging, and it remains unclear how far they meet evidentiary chain-of-custody requirements or legal audit standards.

\textbf{R3.} Flower supports authentication through OpenID Connect and role-based access control through SuperLink\footnote{SuperLink is the server-side component in Flower that orchestrates communication between distributed clients.}, OpenFGA, and OIDC providers. This gives a foundation for defining organizational roles and responsibilities. It does not provide the higher-level governance processes that forensic investigations need, such as inter-agency agreements, liability allocation, or procedural accountability.

\textbf{R4.} Flower integrates FedDebug \citep{FedDebug} and TraceFL \citep{TraceFL}. FedDebug lets a developer inspect training round by round and client by client and identify faulty clients, using differential testing on neuron activations to detect unusual patterns or backdoors. TraceFL attributes a specific prediction of the global model to the clients that contributed to it. Flower also supports interpretable models such as XGBoost-FedXGBllr. All of these tools address developers and researchers. How their output could be turned into an explanation that holds up in court is an open question.

\textbf{R5.} Flower works with PyTorch, TensorFlow, Hugging Face, and scikit-learn, supports Docker and Kubernetes deployment, and provides hook-based extensibility. This interoperability is oriented towards research and production ML environments, and it is unclear how it transfers to forensic toolchains.

\textbf{R6.} Flower does not integrate with forensic evidence management or case workflow systems, so forensic use requires manual integration. Integration with evidence acquisition, chain-of-custody management, or forensic analysis platforms has not been demonstrated.

\textbf{R7.} Flower provides extensive documentation, more than twelve quickstart tutorials, and a command-line interface. Usability for machine learning practitioners does not automatically mean usability for forensic analysts or investigators, who may have no ML background.

%%=====
\bibliographystyle{model5-names}
\bibliography{refs_cited}

\end{document}